\magnification=1200
\settabs 18 \columns

\baselineskip=15 pt
\topinsert \vskip 1.50 in
\endinsert

\def\sqr#1#2{{\vcenter{\vbox{\hrule height.#2pt
 \hbox{\vrule width.#2pt height#1pt \kern#1pt
 \vrule width.#2pt} \hrule height.#2pt}}}}

\def\operp{\hbox{${\kern+.25em{\bigcirc}
\kern-.85em\bot\kern+.85em\kern-.25em}$}}

\def\lsim{\;\raise0.3ex\hbox{$<$\kern-0.75em\raise-1.1ex\hbox{$\sim$}}\;}
\def\gsim{\;\raise0.3ex\hbox{$>$\kern-0.75em\raise-1.1ex\hbox{$\sim$}}\;}
\def\no{\noindent}

\def\ce{\centerline}
\def\ve{\vfill\eject}
\def\rdots{\mathinner{\mkern1mu\raise1pt\vbox{\kern7pt\hbox{.}}\mkern2mu
 \raise4pt\hbox{.}\mkern2mu\raise7pt\hbox{.}\mkern1mu}}

\def\e e{$e^+ e^-$ }



\rightline{UCLA/01/TEP/29}
\rightline{October 2001}
\vskip1.0cm

\ce{{\bf ON $q$-ELECTROWEAK}}
\vskip.5cm

\ce{\it R. J. Finkelstein}
\vskip.3cm
\ce{Department of Physics and Astronomy}
\ce{University of California, Los Angeles, CA  90095-1547}
\vskip1.0cm

\no {\bf Abstract.}  The $q$-electroweak theory obtained by replacing
$SU(2)$ by $SU_q(2)$ in the Weinberg-Salam model is experimentally not
distinguishable from the standard model at the level of the doublet
representation.  However, differences between the two theories should be
observable when higher dimensional representations are taken into account.
In addition the possibility of probing non-local structure may be offered
by the $q$-theory.
\ve

\line{{\bf Introduction.} \hfil}
\vskip.3cm

The Weinberg-Salam model is obtained by gauging $SU(2)_L \times U(1)$, where
$SU(2)_L$ is the chiral isotopic spin group and $U(1)$ is the weak hypercharge
group.$^1$  Since $SU(2)$ is a degenerate form of $SU_q(2)$, it may be
interesting to consider the model obtained by gauging $SU_q(2)_L\times U(1)$.
It is reasonable to do this since $SU(2)$, unlike the Poincar\'e group, is a
phenomenological group and $SU_q(2)$ may also be phenomenologically useful.
In addition $SU_q(2)$ possibly offers a probe of the non-local (solitonic)
structure of massive particles.

Although $SU_q(2)$ is discussed in the mathematical literature as a quasi
triangular Hopf algebra,$^2$ we shall here make use of the usual language
of Lie groups in order to make connections with the standard theory.  We shall
first summarize the necessary information about $SU_q(2)$.
\vskip.5cm

\line{{\bf 2. Irreducible Representations of $SU_q(2)$.} \hfil}
\vskip.3cm

The two-dimensional representation of $SL_q(2)$ may be defined by
$$
T\epsilon T^t = T^t\epsilon T = \epsilon \eqno(2.1)
$$
\no where $t$ means transpose and
$$
\epsilon = \left(\matrix{0 & q_1^{1/2} \cr
-q^{1/2} & 0 \cr} \right) \qquad q_1 = q^{-1} \eqno(2.2)
$$
\no Set
$$
T = \left(\matrix{\alpha & \beta \cr \gamma & \delta \cr} \right)
\eqno(2.3)
$$
\no Then 
$$
\eqalign{\alpha\beta &= q\beta\alpha \cr
\delta\beta &= q_1\beta\delta \cr
\hfil \cr} \qquad
\eqalign{\alpha\gamma &= q\gamma\alpha \cr
\delta\gamma &= q_1\gamma\delta \cr
\hfil \cr} \qquad
\eqalign{&\alpha\delta-q\beta\gamma = 1 \cr
&\delta\alpha-q_1\beta\gamma = 1 \cr
&\beta\gamma = \gamma\beta \cr} \eqno(2.4)
$$
\no If $q=1$, Eqs. (2.4) are satisfied by complex numbers and $T$ is
defined over a continuum, but if $q\not= 1$, then $T$ is defined only
over this algebra--a non-commuting space.

A two-dimensional representation of $SU_q(2)$ may be obtained by going to a
matrix representation of (2.4) and setting
$$
\gamma = -q_1\bar\beta \qquad \delta = \bar\alpha \eqno(2.5)
$$
\no where the bar means Hermitian conjugate.  Then
$$
\eqalign{\alpha\beta &= q\beta\alpha \cr
\alpha\bar\beta &= q\bar\beta\alpha \cr} \qquad
\eqalign{&\alpha\bar\alpha + \beta\bar\beta = 1 \cr
&\bar\alpha\alpha + q_1^2\bar\beta\beta = 1 \cr} \qquad
\beta\bar\beta = \bar\beta\beta \eqno(A)
$$
\no and $T$ is unitary:
$$
\bar T = T^{-1}
$$
\no If $q=1$, $(A)$ may be satisfied by complex numbers and $T$ is a
$SU(2)$ unitary-simplectic matrix.  If $q\not= 1$, there are no finite
representations of $(A)$ unless $q$ is a root of unity.  We shall assume
that $q$ is real and $q<1$.

The irreducible representations of $SU_q(2)$ are as follows:$^3$
$$
{\cal{D}}^j_{mm^\prime}(\alpha,\bar\alpha,\beta,\bar\beta) =
\Delta^j_{mm^\prime}\sum_{s,t}\bigg\langle\matrix{n_+\cr s\cr}\bigg\rangle_1
\bigg\langle\matrix{n_-\cr t\cr}\bigg\rangle_1q_1^{t(n_1-s+1)}
(-)^t\delta(s+t,n_+^\prime)\alpha^s\beta^{n_+-s}\bar\beta^t
\bar\alpha^{n_--t}
$$
\no where
$$
\eqalign{n_\pm &= j\pm m \cr
n_\pm^\prime &= j\pm m^\prime \cr} \qquad
\bigg\langle\matrix{n\cr s\cr}\bigg\rangle_1 = {\langle n\rangle_1!\over
\langle s\rangle_1!\langle n-s\rangle_1!} \qquad
\langle n\rangle_1 = {q_1^{2n}-1\over q_1^2-1} 
$$
$$
\Delta^j_{mm^\prime} = \biggl[{\langle n_+^\prime\rangle!
\langle n^\prime_-\rangle!\over \langle n_+\rangle!\langle n_-\rangle!}\biggr]^{1/2} \qquad q_1 = q^{-1}
$$
\no In the limit $q=1~{\cal{D}}^j_{mm^\prime}$ become the Wigner functions,
$D^j_{mm^\prime}(\alpha,\beta,\gamma)$, the irreducible representations of
$SU(2)$.
\vskip.5cm

\line{{\bf 3. The Lie Algebra of $SU_q(2).^{~3}$} \hfil}
\vskip.3cm

The Lie algebra of $SU_q(2)$ may be obtained in the following way.  The
two-dimensional representation $T$ may be Borel factored:
$$
T = \left(\matrix{\alpha & \beta \cr \gamma & \delta} \right) =
e^{\beta\sigma_+}e^{\lambda\theta\sigma_3}e^{C\sigma_-} \eqno(3.1)
$$
\no The algebra of $(\alpha,\beta,\gamma,\delta)$ is then inherited by
$(B,C,\theta)$ as
$$
\eqalign{&(B,C) = 0 \quad (\theta,B) = B \quad (\theta,C) = C \cr
&\lambda = \ln q \cr} \eqno(3.2)
$$
\no The $2j+1$ dimensional irreducible representation of $SU_q(2)$ shown in
(2.6) may be rewritten in terrms of $(B,C,\theta)$.  Then by expanding to
terms linear in $(B,C,\theta)$ one has
$$
{\cal{D}}^j_{mm^\prime}(B,C,\theta) = {\cal{D}}^j_{mm^\prime}(0,0,0) +
B(J^j_B)_{mm^\prime} + C(J^j_C)_{mm^\prime} + 2\lambda\theta
(J^j_\theta)_{mm^\prime} + \ldots \eqno(3.3)
$$
\no where the non-vanishing matrix coefficients $(J_B^j)_{mm^\prime}$,
$(J^j_C)_{mm^\prime}$, and $(J^j_\theta)_{mm^\prime}$ are
$$
\eqalignno{\langle m-1|J_B^j|m\rangle &= (\langle j+m\rangle_1
\langle j-m+1\rangle_1)^{1/2} & (3.4B) \cr
\langle m+1|J_C^j|m\rangle &= (\langle j-m\rangle_1
(\langle j-m\rangle_1\langle j+m+1\rangle)^{1/2} & (3.4C) \cr
\langle m|J_\theta^j|m\rangle &= m & (3.4\theta) \cr}
$$
\no and where
$$
\langle n\rangle_1 = \langle n\rangle_{q_1^2} = {q_1^{2n}-1\over q_1^2-1}
\eqno(3.5)
$$
\no is a basic integer corresponding to $n$.  The $(B,C,\theta)$ and
$(J_B,J_C,J_\theta)$ are generators of two dual alglebras satisfying the
following commutation rules:
$$
(J_B,J_\theta) = -J_B \quad (J_C,J_\theta) = J_C \quad
(J_B,J_C) = q_1^{2j-1}[2J_\theta] \eqno(3.6)
$$
$$
(B,C) = 0 \qquad (\theta,B) = B \qquad (\theta,C) = C \eqno(3.7)
$$
\no Here
$$
[x] = {q^x-q_1^x\over q-q_1} \eqno(3.8)
$$
\vskip.5cm

\line{{\bf 4. Fundamental and Adjoint Representations.} \hfil}
\vskip.3cm

For comparison with the standard theory we need to know the fundamental and
adjoint representations.  In the fundamental representation $(j=1/2)$
we have by (3.4)-(3.6):
$$
J_B = \left(\matrix{0 & 1 \cr 0 & 0 \cr}\right) \quad
J_C = \left(\matrix{ 0 & 0 \cr 1 & 0 \cr}\right) \quad
J_\theta = \left(\matrix{1/2 & 0 \cr 0 & -1/2 \cr}\right) \eqno(4.1)
$$
\no and
$$
(J_B,J_C) = 2J_\theta \quad (J_B,J_\theta) = -J_B \quad
(J_C,J_\theta) = J_C \eqno(4.2)
$$
\no In the adjoint representation $(J=1)$ we have by the same relations
(3.4)-(3.6):
$$
J_B = \left(\matrix{0 & x & 0 \cr 0 & 0 & x \cr 0 & 0 & 0 \cr}\right) \quad
J_C = \left(\matrix{0 & 0 & 0 \cr x & 0 & 0 \cr 0 & x & 0 \cr}\right) \quad
J_\theta = \left(\matrix{1 & \hfil & \hfil \cr \hfil & 0 & \hfil \cr
\hfil & \hfil & -1 \cr}\right)
\eqno(4.3)
$$
\no and
$$
(J_B,J_C) = x^2J_B \quad (J_B,J_\theta) = -J_B \quad
(J_C,J_\theta) = J_C \eqno(4.4)
$$
\no where
$$
x = \langle 1|J_B|0\rangle = \langle 2\rangle_1^{1/2}
\langle 1\rangle_1^{1/2} \eqno(4.5)
$$
\no Here $\langle n\rangle_1$ is given by (3.5).  Then
$$
x = \langle 2\rangle_1^{1/2} = (1+q_1^2)^{1/2} \eqno(4.6)
$$
\no and
$$
(J_B,J_C) = (1+q_1^2)J_\theta \eqno(4.7)
$$
\no In general we have by (3.4) for all representations:
$$
(J_B,J_C) = q_1^{2j-1}~[2J_\theta]_q \eqno(4.8)
$$
\no where
$$
\eqalign{[2J_\theta]_q &= {q^{2J_\theta}-q_1^{2J_\theta}\over
q-q_1} \cr
&= {e^{2\lambda J_o}-e^{-2\lambda J_o}\over q-q_1} \cr} \eqno(4.9)
$$
\no and
$$
q = e^\lambda \eqno(4.10)
$$
\no Then
$$
\eqalign{[2J_\theta]_q &= {2\sinh 2\lambda J_\theta\over q-q_1} \cr
&= {2\over q-q_1}\biggl[2\lambda J_\theta +{(2\lambda J_\theta)^3\over 3!}
+ \ldots \biggr] \cr} \eqno(4.12)
$$
\no In the fundamental representation $J_\theta$ is given by (4.1) and
$$
J_\theta^2 = {1\over 4} I \eqno(4.13)
$$
\no Then by (4.12)
$$
\eqalignno{[2J_\theta]_q &= {2J_\theta\over q-q_1}
\biggl[2\lambda + {2\lambda^3\over 3!} + \ldots \biggr] & (4.14) \cr
[2J_\theta]_q &= {4J_\theta\over q-q_1} \sinh\lambda \cr
&= 2J_\theta & (4.15) \cr}
$$
\no Then one sees that (4.8), with the aid of (4.15) reduces to (4.2) in
the fundamental representation.

In the adjoint representation $J_\theta$ is given by (4.3) and
$$
J_\theta^3 = J_\theta \eqno(4.16)
$$
\no Hence every odd power of $J_\theta$ may be reduced as follows:
$$
J_\theta^{2p+1} = J_\theta^{3p-p+1} = J_\theta^{p-p+1} = J_\theta
\eqno(4.17)
$$
\no By (4.12) and (4.17)
$$
\eqalign{[2J_\theta]_q &= {2J_\theta\over q-q_1}
\biggl[2\lambda + {(2\lambda)^3\over 3!} + \ldots \biggr] \cr
&= {2J_\theta\over q-q_1} sinh2\lambda \cr
&= J_\theta {q^2-q_1^2\over q-q_1} \cr
&= (q+q_1) J_\theta \cr}  \eqno(4.18)
$$
\no Hence by (4.8) in the $J=1$ representation
$$
\eqalign{(J_B,J_C) &= q_1(q+q_1) J_\theta \cr
&= (1+q_1^2)J_\theta \cr} \eqno(4.19)
$$
\no in agreement with (4.7).

Let us now change to the $(J_1,J_2,J_3)$ basis as follows:
$$
\eqalign{J_B &= J_1 + i~J_2 \cr
J_C &= J_1 -i~J_2 \cr
J_\theta &= J_3 \cr} \eqno(4.20)
$$
\no In the fundamental representation, Eqn. (4.2) may be rewritten by
(4.20) as
$$
\eqalign{(J_1,J_2) &= i~J_3 \cr
(J_2,J_3) &= i~J_1 \cr
(J_3,J_1) &= i~J_2 \cr} \eqno(4.21a)
$$
\no where as usual
$$
J_1 = {1\over 2} \left(\matrix{0 & 1 \cr 1 & 0 \cr} \right) \qquad
J_2 = {1\over 2} \left(\matrix{0 & -i \cr i & 0 \cr} \right) \qquad
J_3 = {1\over 2}\left(\matrix{i & 0 \cr 0 & -1 \cr} \right) 
\eqno(4.21b)
$$
\no In the adjoint representation Eqs. (4.4) may be rewritten by (4.20) as
$$
\eqalign{(J_1,J_2) &= i{\langle 2\rangle\over 2} J_3 \cr
(J_2,J_3) &= i~J_1 \cr
(J_3,J_1) &= i~J_2 \cr} \eqno (4.22a)
$$
\no where
$$
J_1 = {\langle 2\rangle_1^{1/2}\over 2}
\left(\matrix{0 & 1 & 0 \cr 1 & 0 & 1 \cr 0 & 1 & 0 \cr}
\right) \quad
J_2 = {\langle 2\rangle_1^{1/2}\over 2}
\left(\matrix{0 & -i & 0 \cr i & 0 & -i \cr 0 & i & 0 \cr}\right) \quad
J_3 = \left(\matrix{1 & \hfil & \hfil \cr \hfil & 0 & \hfil \cr
\hfil & \hfil & -1 \cr}\right) \eqno(4.22b) 
$$
\no Now write (4.21a) and (4.22a) in the following familiar way:
$$
(t_a,t_b) = f_{ab}^{~~m} t_m \eqno(4.23)
$$
\no The $f_{ab}^{~~m}$ are the same for all representations of $SU(2)$.
For $SU_q(2)$ however, the $f_{ab}^{~~m}$ are different in the fundamental
and adjoint representations, and in higher representations the commutator
is no longer linear in the generators.

When (4.23) does hold, however, we also have
$$
{\rm Tr}(t_a,t_b)t_d = f_{ab}^{~~m}~{\rm Tr}~t_mt_d \eqno(4.24)
$$
\no Introducing the ``group metric", namely
$$
g_{md} = {\rm Tr}~t_mt_d \eqno(4.25)
$$
\no we have
$$
\eqalignno{f_{abd} &\equiv f_{ab}^{~~m}g_{md} = {\rm Tr}(t_a,t_b)t_d 
& (4.26) \cr
&= {\rm Tr}(t_d,t_a)t_b = -{\rm Tr}(t_a,t_d)t_b \cr
&= f_{dab} = -f_{adb} & (4.27) \cr}
$$
\no Then the $f_{abd}$ are the completely antisymmetric structure constants.
For the adjoint representation of $SU_q(2)$ we have by (4.22b)
$$
\eqalignno{g_{ab} &= g_a\delta_{ab} & (4.28a) \cr
g_1 &= g_2 = \langle 2\rangle_1 & (4.28b) \cr
~~~~ g_3 &= 2 & (4.28c) \cr}
$$
\no The ``structure constants" exhibited by (4.22) are
$$
\eqalign{f_{12}^{~~3} &= i{\langle 2\rangle\over 2} \cr
f_{23}^{~~1} &= f_{31}^{~~2} = i \cr} \eqno(4.29)
$$
\no Then by (4.26), (4.28), and (4.29)
$$
f_{123} = f_{231} = f_{312} = i\langle 2\rangle \epsilon_{123} \eqno(4.30)
$$
\no In the light of these properties of $SU_q(2)$ let us now consider
$q$-electroweak.
\vskip.5cm

\line{{\bf 5. $q$-Electroweak.} \hfil}
\vskip.3cm

In the Weinberg-Salam model the Lagrangian density is$^4$
$$
\eqalign{{\cal{L}} = &- {1\over 4}(G^{\mu\nu}G_{\mu\nu} + H^{\mu\nu}H_{\mu\nu}) +i(\bar LD\!\!\!/L + \bar RD\!\!\!/R) \cr
&+(\overline{D\phi})(D\phi)-V(\bar\varphi\varphi) - {m\over p_o}
(\bar L\varphi R + \bar R\bar\varphi L) \cr} \eqno(5.1)
$$
\no where the covariant derivative is
$$
D = \partial + ig\vec W\vec t + ig^\prime W_ot_o \eqno(5.2)
$$
\no Here $\vec W^\mu$ and $W_o^\mu$ are the connection fields of
$SU(2)_L$ and $U(1)$, the chiral isotopic spin and hypercharge groups with
independent coupling constants $g$ and $g^\prime$, while $G$ and $H$ are the
corresponding field strengths.  The Lagrangian (5.1) also contains the
contribution of one lepton doublet and the mass generating Hibbs doublet
$\varphi$.

In (5.2), the expression for the covariant derivative, the matrices $\vec t$
and $t_o$ are the generators of the $SU(2)$ and $U(1)$ groups.  If we now
pass to $SU_q(2)$ without changing $U(1)$, Eqs. (5.2) will be unchanged since
Eqs. (4.1) and (4.2) hold for both $SU(2)$ and $SU_q(2)$.

If no other changes are made in the standard theory, the expression for the
covariant derivative may be rewritten as follows:
$$
D = \partial + ig\biggl[{1\over 2} W_-t_+ + {1\over 2} W_+t_-\biggr] + ie
[QA + Q^\prime{\bf Z}] \eqno(5.3)
$$
\no where $A$ and $Z$ are the Maxwell and Weinberg-Salam neutral fields while
$W_+$ and $W_-$ are the charged vector fields and where by Weinberg-Salam:
$$
\eqalignno{Q &= t_3+t_o & (5.4) \cr
Q^\prime &= t_3\cot\theta-t_o\tan\theta & (5.5) \cr
e &= g\sin\theta = g^\prime\cos\theta & (5.6) \cr}
$$
\no The masses of the charged $W^\pm$ and neutral {\bf Z} are then determined
by the Higgs mechanism as implemented in the unitary gauge by (5.1) and 
(5.3) to be
$$
\eqalignno{M^2_W &= {1\over 2} g^2p_o^2 & (5.7) \cr
M_Z &= {M_W\over\cos\theta} & (5.8) \cr}
$$
Eq. (5.8) has been checked by direct and independent measurements of the
masses of the $W^\pm$ and {\bf Z} on the one hand and of the Weinberg angle
$\theta$ on the other.

None of these relations is changed on passing from the algebra of $SU(2)$ to the
algebra of $SU_q(2)$ since the three $\vec t$ matrices in the fundamental
representation are the same for both.  In particular (5.8) still holds and
provides strong experimental confirmation of the Weinberg-Salam model.
On the other hand higher dimensional representations, including already the
adjoint representation, will be different for $SU_q(2)$ as one sees from
(4.3) and (4.4) and also from the Clebsch-Gordan coefficients of $SU_q(2).^{~5}$
Hence differences introduced by the $q$-theory are in principle detectable,
although they would be difficult to quantitatively isolate from radiative
corrections that are also present.  Let us next see how differences in the
adjoint representation manifest themselves in a more general $q$-gauge theory.
\vskip.5cm

\line{{\bf 6. General $q$-Gauge Theory.} \hfil}
\vskip.3cm

The Lagrangian density for the standard non-Abelian gauge theory may be
written in the following form:
$$
{\cal{L}} = -{1\over 4} g_{ab} F^a_{\mu\nu} F^{b\mu\nu} + {\cal{L}}_m(\psi,
\nabla_\nu\psi) \eqno(6.1)
$$
\no where
$$
\eqalignno{\nabla_\mu &= \partial_\mu + A_\mu & (6.2) \cr
A_\mu &= ig A^a_{~\mu}t_a & (6.3) \cr
F_{\mu\nu} &= (\nabla_\mu,\nabla_\nu) & (6.4) \cr}
$$
\no Here $\nabla_\mu$ is the covariant derivative.  In (6.3) $g$ is the weak
coupling constant and $t_a$ are the generators of $SU(2)$ in the adjoint
representation.  ${\cal{L}}$ is invariant under infinitesimal gauge
transformations as follows:
$$
\left(\matrix{\delta F^a \cr \delta\nabla^a \cr \delta\varphi^a \cr
\delta(\nabla\varphi)^a \cr} \right) = i~\epsilon^cf^a_{~cb}
\left(\matrix{F^b \cr \nabla^b \cr \phi^b \cr (\nabla\phi)^b \cr}\right)
\eqno(6.5)
$$
\no We may pass from these relations for the standard theory to the
corresponding relations for the $q$-standard theory by replacing the
$\vec t$ of $SU(2)$ by the $\vec t$ of $SU_q(2)$ in the adjoint representation,
and by replacing the structure constants of the $SU(2)$ algebra by the
corresponding $f^a_{~cb}$ of $SU_q(2)$.

The invariance of scalar products like $g_{ab}F^aF^b$ with the transformation
rule (6.5) still holds for $SU_q(2)$ since
$$
\delta(g_{ab}F^aF^b) \sim \epsilon^cf_{ace}F^aF^e = 0 \eqno(6.6)
$$
\no where $f_{ace}$ is the completely antisymmetric symbol introduced in
(4.26).

Let us now consider the self-coupling of the vector field.  We have
$$
F^a_{\mu\nu} = \partial_\mu A^a_\nu-\partial_\nu A^a_\mu + f^a_{~bc}
A^b_\mu A^c_\nu \eqno(6.7)
$$
\no and
$$
-{1\over 4} g_{mn}F^m_{\mu\nu}F^{n\mu\nu} = -{1\over 4} g_{mn}(\partial_\mu
A^m_\nu-\partial_\nu A^m_\mu + f^m_{bc}A^b_\mu A^c_\nu)
(\partial^\mu A^{\nu n}-\partial^\nu A^{\nu n} + f^n_{k\ell}A^{k\mu}A^{\ell\nu})
\eqno(6.8)
$$
\no The trilinear couplings are then
$$
\eqalign{\sim &g_{mn}f^m_{bc}A^b_\mu A^c_\nu(\partial^\mu A^{\nu n}
-\partial^\nu A^{\mu n}) \cr
&= f_{nbc}A^b_\mu A^c_\nu(\partial^\mu A^{\nu n}-\partial^\nu A^{\mu n}) \cr}
\eqno(6.9)
$$
\no where
$$
\eqalign{f_{nbc} &= i(\langle 2\rangle_1\epsilon_{nbc} \quad
{\rm by ~ (4.30)} \cr
&= i(1+q_1^2)\epsilon_{nbc} \cr} \eqno(6.10)
$$
\no The weak coupling constant associated with the $SU_q(2)$ theory is
therefore greater than the corresponding coupling constant associated with
the $SU(2)$ theory by the ratio ${1\over 2}(1+q_1^2)$, when $q_1>1$.  Since
the factor $(1+q_1^2)$ can be absorbed into a new definition of the
coupling constant, however, there is no observable difference between the
two theories coming from the trilinear couplings.

The quadlinear couplings are on the other hand
$$
\sim g_{mn} f^m_{bc} f^n_{k\ell} A^b_\mu A^c_\nu A^{k\mu}A^{\ell\nu}
\eqno(6.11)
$$
\no Here
$$
g_{mn}f^m_{bc}f^n_{k\ell} = f_{nbc}f^n_{k\ell} \eqno(6.12)
$$
\no At this point there appears a real difference between the $SU(2)$ and
$SU_q(2)$ theories coming from the $f^n_{k\ell}$ as shown by (4.29).  This difference arises
from the asymmetry in (4.22).  It distinguishes one preferred direction in
isotopic spin space, and in principle should be experimentally detectable.

Let us next consider the new degrees of freedom provided by the $q$-theories.
\ve

\line{{\bf 7. The Dual Algebra.} \hfil}
\vskip.3cm

There are two $q$-algebras, $G_q$ and $g_q$, that are respectively 
deformations of the Lie group $(G)$ and its algebra $(g)$.  The matrix
elements of the representation matrices of $G$, $g$, and $g_q$ commute,
but the same is not true for $G_q$ as one sees in (2.4) for example.

The deformation of the standard model that we have been discussing is based
on $g_q$, the deformed Lie algebra and we have seen that it is a possibly
acceptable modification of the standard model.

There is also a more speculative construction based on $G_q$, the deformed
Lie group described in Section 2.  In this construction we assume that all
fields lie in the algebra of $SU_q(2)$.  Thus a generic field may be expanded
as follows:
$$
\eqalignno{\psi(x) &= \sum \varphi^j_{mn}{\cal{D}}^j_{mn}(\alpha|q)
& (7.1) \cr
\bar\psi(x) &= \sum \bar\varphi^j_{mn}{\cal{D}}^j_{mn}(\alpha|q) & (7.2) \cr}
$$
\no where the ${\cal{D}}^j_{mn}(\alpha|q)$ are the irreducible representations
of $SU_q(2)$.

The partial fields $\varphi^j_{mn}(x)$ appearing in (7.1) may be expanded
in Fock annihilation and creation operators
$$
\varphi^j_{mn}(x) = {1\over (2\pi)^{3/2}}\int {d\vec p\over (2p_o)^{1/2}}
\bigl[e^{-ipx}a^j_{mn}(\vec p) + e^{ipx}\bar a^j_{mn}(\vec p)\bigr]
\eqno (7.3)
$$
\no To illustrate, consider the normal ordered Hamiltonian of a global
scalar field
$$
H^{(q)} = {1\over 2} h\int :\biggl[\sum^3_0 \partial_k\bar\psi\partial_k\psi
+m_o^2\bar\psi\psi\biggr]:d\vec x \eqno(7.4)
$$
\no where the symbol $h$ standing before the spatial integral denotes
Woronowitz integration$^6$ over the algebra $G_q$.  Then by (7.1)-(7.3)
$$
H^{(q)} = \int d\vec p p_o\sum_{\scriptstyle jmn\atop\scriptstyle 
j^\prime m^\prime n^\prime} h(\bar{\cal{D}}^j_{mn}{\cal{D}}^{j^\prime}_{m^\prime n^\prime}){1\over 2}:\bigl[\bar a^j_{mn}(p)a^{j^\prime}_{m^\prime n^\prime}(p)
+ a^{j^\prime}_{m^\prime n^\prime}(p) \bar a^j_{mn}(p)\bigr]:
\eqno(7.5)
$$
\no The orthogonality of the $q$-irreducible representations is now expressed
in terms of the following integral over the algebra:$^6$
$$
h(\bar{\cal{D}}^j_{mn}{\cal{D}}^{j^\prime}_{m^\prime n^\prime}) =
\delta^{jj^\prime}\delta_{mm^\prime}\delta_{nn^\prime}
{q^{2n}\over [2j+1]_q} \eqno(7.6)
$$
\no Hence
$$
H^{(q)}|N(p);jmn\rangle = {p_oq^{2n}\over [2j+1]_q}|N(p);jmn\rangle \eqno(7.7)
$$
\no and the mass of a single field particle with ``internal" quantum numbers
$(jmn)$ is$^7$
$$
{m_oq^{2n}\over[2j+1]_q}
$$
\no The spectrum resembles the square root of the spectrum of the $q$-H atom.

One may regard the mass $m_o$ in (7.4) as generated by a Higgs type mechanism.
Then by (7.8) the field particles have excited states that depend on the
internal quantum numbers.  The existence of excited states, impossible for a
point particle, reveals non-local or solitonic structure.  In the scenario
sketched in this note, the dual algebras provide complementary pictures of
the field particles: the $G_q$ picture is microscopic and describes
solitonic structure, while the $g_q$ picture is phenomenological and 
approximates the standard theory.  One may also speculate that $SU(3)$
flavor and $SU(3)$ color may be similarly related: that they may be based
on the dual algebras of $SU_q(3)$.
\vskip.5cm

\line{{\bf References.} \hfil}
\vskip.3cm

\item{1.} S. Weinberg, {\it The Quantum Theory of Fields}, Vol. 2, Cambridge
(1996).
\item{2.} J. Fuchs, see for example {\it Affine Lie Algebras and Quantum
Groups}, Cambridge (1992).
\item{3.} See for example, R. Finkelstein, Lett. Math. Phys. {\bf 29}, 75
(1993).
\item{4.} See for example, K. Huang, {\it Quarks, Leptons, and Gauge Fields},
World Scientific (1982).
\item{5.} See for example, C. Cadavid and R. Finkelstein, J. Math. Phys.
{\bf 36}, 1912 (1995).
\item{6.} S. Woronowicz, Commun. Math. Phys. {\bf 111}, 613 (1987).
\item{7.} R. Finkelstein, hep-th/0106283.

\end